\documentstyle[preprint,aps]{revtex}
\begin{document}

\draft

\title{Normal State Property of the t-J Model}

\author{Yu-Liang Liu}
\address{International Center for Theoretical Physics, P. O. Box 586, 34100
 Trieste, Italy}

\maketitle

\begin{abstract}
Using the spin-hole coherent state representation and taking a long range 
antiferromagnetic N\`{e}el order as a background of the localized spin 
degree part, we have studied the normal state behavior of the t-J model, 
and shown that a strongly short-range antiferromagnetic correlation of 
the localized spin degree part is responsible for the anomalous 
non-Korringa-like relaxation behavior of the planar copper spin, the 
Korringa-like behavior of the planar oxygen spin may derive from the 
charge degree part describing a Zhang-Rice spin-singlet; The charge 
degree part feels a strongly staggered magnetic field induced by this 
short-range antiferromagnetic correlation as a doping hole 
hopping, this staggered magnetic field enforces the charge degrees to 
have different responses to external magnetic and electric fields and to 
show two relaxation rate behaviors corresponding to the planar 
resistivity and Hall angle, respectively. We have found that the 
temperature dependence of magnetoresistance is $T^{-n}$, $n\simeq 3$, 
near the optimal doping, $n\simeq 4$, in the underdoping region, 
violating Kohler's rule, the transport relaxation rate is of the order 
of $2k_{B}T$, all that are 
consistent with the normal state of the cuprate superconductors.

\end{abstract}
\vspace{1cm}


\newpage

\section{Introduction}

Recently, the significant progress has been made in the understanding of 
the low energy spin dynamics of the normal state of the cuprate 
superconducting materials in both theoretical\cite{1}-\cite{4b} and 
experimental\cite{5}-\cite{7b} aspects. In the undoping case, the spin 
dynamics 
of the cuprates, such as $La_{2}CuO_{4}$, is well described by the 
quantum Heisenberg model on a square lattice of $Cu$ sites. The authors 
of Refs.[1,2] have extensively studied it by using the scalling and 
renormalization group theory and /or large-N expansion methods, and have 
given some valuable results which are in good agreement with the current 
experimental data. However, in the doped case, up to now there is not a 
general consensus on choosing a microscopic theory qualitatively to 
describe the unusually magnetic and transport properties of the normal 
state over the entire doping range from insulator to high doped 
compounds, although many models have been proposed to describe them.

In all hole-type cuprates near optimal doping, the mostest important 
properties of the normal state are that: a). The linear dependence of the 
in-plane resistivity $\rho$ on temperature (T) has been confirmed from 
$T_{c}$ up to temperature as high as 1000K\cite{7c}. b). The in-plane Hall 
resistivity $\rho_{xy}=R_{H}B$ varies strongly over a wide range of T ($B$, 
external magnetic field; $R_{H}$, the Hall coefficient), it falls as 
$\sim T^{-1}$ between $T_{c}$ and temperature as high as 500K. 
Anderson\cite{7d} prodicted that the cotangent of the Hall angle should 
vary with impurity content $n_{i}$ as $cot\theta_{H}=\alpha T^{2}+\beta 
n_{i}$, which was immediately confirmed by Ong\cite{7e} and 
others\cite{7f}\cite{7g}. c). The magnetoresistance 
$\Delta\rho/\rho\propto B^{2}T^{-n},\; n\simeq 3\sim 4$, strongly 
violates Kohler's rule\cite{7i}\cite{7j}. d). The relaxation behaviors at 
various sites show 
sharply contrast. The relaxation behavior of the planar oxygen obeys 
Korringa-like behavior, while the planar copper sharply shows 
non-Korringa behavior. e). Dynamical antiferromagnetic correlations 
persist in all the metallic state and the superconducting state. These 
properties provide significant constraints on candidate descriptions of 
their anomalous normal state behavior. The items a),b),c) strongly indicate 
that there exist two different relaxation times in the system\cite{7d} 
$1/\tau_{tr}\sim T$, and $1/\tau_{H}\sim T^{2}$. While the items d),e) 
strongly show that among the planar copper spins there exists a strongly 
commensurate antiferromagnetic correlation, the planar oxygen, residing 
at the middle point between two nearest neighbor coppers, is not affected 
by this commensurate antiferromagnetic correlation and shows the 
Korringa-like relaxation behavior, at least it is true for the 
$YBa_{2}Cu_{3}O_{6+x}$ materials. For the $La_{2-x}Sr_{x}CuO_{4}$ system, 
although the neutron scattering experiments show at low temperature four 
incommensurate peaks in the spin fluctuation spectrum, whose position 
depends on the level of $Sr$ doping\cite{7l}, the nuclear magnetic 
resonance experiment shows that the property d) is remaining 
invariance\cite{7b}.

The unusually physical properties of the normal state of the cuprate 
superconducting materials may originate from their strongly 
antiferromagnetic correlation. The doping will destroy the long range 
antiferromagnetic correlation, but the system still maintains a strongly 
short range antiferromagnetic correlation. In Refs. 4-6, we have given a 
detail study following this idea, and obtained some results which can 
qualitatively explain the unusually physical properties of the normal 
state. In this paper, using the similar method as in Refs. 4-6, we study the 
normal state behavior of the t-J model. The t-J model carries on the 
important electronic strongly correlated property of the cuprate 
superconducting materials, through completing studying its property we 
hope to get more understanding of the strongly correlated electronic 
system and the cuprate superconducting materials. It is well-known that the 
gauge 
theory of the t-J model\cite{8}-\cite{10} gives a better description  
to the temperature dependence of the in-plane resistivity of the normal 
state, but up to now one has not known 
whether it can also give a reasonable description to the temperature 
dependence of the Hall coefficient and the magnetic behavior of the normal 
state, i.e., can it show not only the two relaxation time behaviors but 
also the strongly antifrromagnetic correlation behavior? In this paper, 
we show that the unusual magnetic behavior of the normal state 
is induced by a strongly short-range antiferromagnetic correlation among 
the localized spin degrees on the copper sites, while because of there 
existing this strongly 
short-range antiferromagnetic correlation in the localized spin degree 
part, the charge degree part will feel a strong staggered magnetic field 
as the doping hole hopping, this staggered magnetic field drastically 
influences the behavior of the charge degree part, and enforces the 
charge degree part to have different responses to external magnetic field 
and electric field and to show two relaxation time behaviors 
corresponding to the in-plane resistivity and Hall angle, respectively.
In the usual slave boson(or fermion) description of the t-J model, the 
spin degree and the charge degree of electrons are separated, the spin 
degree part effectively describes the localized spins on the copper 
sites, while the charge degree part effectively describes the 
spin-singlet (or Zhang-Rice singlet) of the doping hole spin and the copper 
spin\cite{10a}.
However, because the strongly dynamical antiferromanetic correlations 
persist in all the metal state, it is reasonable that we use the 
long range antiferromagnetic N\`{e}el order of the localized spin degerees 
as our starting point to study the normal state behavior of the t-J model.

\section{Spin-hole coherent state representation and spin-charge separation}

We adopt an usual method to deal with the single occupation 
condition by introducing a slave fermion, so the Hamiltonian of the t-J 
model can be written as in a hole representation
\begin{equation}\begin{array}{rl}
H= & t\displaystyle{\sum_{<ij>}}(f^{+}_{j}f_{i}b^{+}_{i\sigma}b_{j\sigma}
+h.c)\\
& +J\displaystyle{\sum_{<ij>}}(1-f^{+}_{i}f_{i})\hat{S}_{i}\cdot\hat{S}_{j}
(1-f^{+}_{j}f_{j})+\displaystyle{\sum_{i}}\lambda_{i}(1-f^{+}_{i}f_{i}
-b^{+}_{i\sigma}b_{i\sigma})
\end{array}\end{equation}
where $\hat{S}_{i}=\frac{1}{2}b^{+}_{i\alpha}\hat{\sigma}_{\alpha\beta}
b_{i\beta}$, $b_{i\sigma}$ is a hard-core boson operator which describes 
the spin degree of the electron, and $f_{i}$ is a fermion operator which 
describes the charge degree of the electron. The electron operator is 
$c_{i\sigma}=f^{+}_{i}b_{i\sigma}$, $\lambda_{i}$ is a Lagrangian 
multiplier which ensures the single occupation condition of the 
electrons. In the spin-hole coherent state representation introduced by 
Auerbach\cite{11}
\begin{equation}
|\hat{\Omega}, \xi>_{S}\equiv |\hat{\Omega}>_{S}\otimes |0>_{f}+
|\hat{\Omega}>_{S-\frac{1}{2}}\otimes\xi f^{+}|0>_{f}
\end{equation}
where $|\hat{\Omega}>_{S}$ is a spin coherent state\cite{12} and $\xi$ is 
an anticommuting Grassmann variable, the partition functional of the 
Hamiltonian (1) can be written as
\begin{equation}
Z=\int D\hat{\Omega}D\xi^{*}D\xi exp\{-\int^{\beta}_{0}[{\it L}_{\Omega}+
{\it L}_{\xi}]\}
\end{equation}
\begin{equation}
{\it L}_{\Omega}=-i\sum_{i}2S\omega_{i}+JS^{2}\sum_{<ij>}(1-\xi^{*}_{i}
\xi_{i})\hat{\Omega}_{i}\cdot\hat{\Omega}_{j}(1-\xi^{*}_{j}\xi_{j})
\end{equation}
\begin{equation}
{\it L}_{\xi}=\sum_{i}\xi^{*}_{i}(\partial_{\tau}+i\omega_{i}+\mu_{i})
\xi_{i}+\sqrt{2}tS\sum_{<ij>}(\xi^{*}_{j}\xi_{i}e^{i\gamma_{ij}}\sqrt{
1+\hat{\Omega}_{i}\cdot\hat{\Omega}_{j}}+h.c)
\end{equation}
where the Berry phase $\omega$ is a functional of the spin order 
parameter $\hat{\Omega}(\tau)$. It is ambiguous modulo $4\pi$, and its 
functional derivative is quite well-behaved\cite{12}
\begin{equation}
\int d\tau\delta\omega=\int d \tau\hat{\Omega}\cdot(\partial_{\tau}
\hat{\Omega}\times\delta\hat{\Omega})
\end{equation}
The parameter $\mu_{i}$ is a chemical potential of the slave fermion 
$\xi$, $\gamma_{ij}$ is the phase factor of 
$_{S}<\hat{\Omega}|b^{+}_{i\sigma}b_{j\sigma}|\hat{\Omega}>_{S}$.
The Lagrangian ${\it L}_{\xi}$ is invariant under following gauge 
transformations
\begin{equation}
\xi_{i}\rightarrow\xi_{i}e^{i\theta_{i}},\;\;
\gamma_{ij}\rightarrow\gamma_{ij}-\theta_{i}+\theta_{j},\;\;
\mu_{i}\rightarrow\mu_{i}+i\partial_{\tau}\theta_{i}
\end{equation}
which derives from the slave fermion representation of the electron 
operator $c_{i\sigma}=f^{+}_{i}b_{i\sigma}$. The single occupation 
condition in (1) disappears in (4) and (5), because in the spin-hole 
coherent state representation the term 
$(1-f^{+}_{i}f_{i}-b^{+}_{i\sigma}b_{i\sigma})$ is equal to zero at each 
site. From the equations (4) and (5), we see that the Lagrangian ${\it 
L}_{\Omega}$ dominates the antiferromagnetic behavior of the system, then 
the Lagrangian ${\it L}_{\xi}$ dominates the ferromagnetic behavior (or 
destroys the antiferromagnetic behavior) of the system because the 
factor $\sqrt{1+\hat{\Omega}_{i}\cdot\hat{\Omega}_{j}}$ is zero for 
antiferromagnetic order and is biggest for ferromagnetic order. According 
to the current experimental data of the cuprate superconducting 
materials, almost all of them show a strongly short range 
antiferromagnetic behavior in the normal state, even in the 
superconducting state, the short range antiferromagnetic behavior also 
appears. Therefore, according to this fact, we take a long range 
antiferromagnetic N\'{e}el order as a background of the spin order parameter
\begin{equation}
\hbar S\hat{\Omega}_{i}\simeq\hbar\eta_{i}\hat{\Omega}(x_{i})
+a^{2}\hat{L}(x_{i})
\end{equation}
where $a^{2}$ is the unit cell volume, $\hat{\Omega}(x_{i})$ is the 
slowly varying N\'{e}el unit vector order, i.e., spin parameter field 
$|\hat{\Omega}(x_{i})|=1$, and $\hat{L}(x_{i})$ is the slowly varying 
magnetization density field, $\hat{\Omega}(x_{i})\cdot\hat{L}(x_{i})=0$. 
The Berry phase term may be separated into two parts
\begin{equation}
S\sum_{i}\omega_{i}\simeq S\sum_{i}\eta_{i}\omega(x_{i})+
\frac{1}{\hbar}\int 
d^{2}x\hat{\Omega}\cdot(\frac{\partial\hat{\Omega}}
{\partial\tau}\times\hat{L})
\end{equation}
where $\omega(x)$ is the solid angle subtended on the unit sphere by the 
closed curve $\hat{\Omega}(x,\tau)$ (parametrized by $\tau$). Because of 
in the long range antiferromagnetic N\'{e}el order approximation, the 
electron hoping must be accompanied with a $\pi-phase$ rotation in spin 
space to match with the nextest neighbor spin orientations, so the t-term 
in (1) must be changed as
\begin{equation}\begin{array}{rl}
f^{+}_{i}f_{j}b^{+}_{j\sigma}b_{i\sigma}=&
\displaystyle{e^{-2i\displaystyle{\sum_{l\neq 
i,j}\theta_{ij}(l)S^{z}_{l}}}}f^{+}_{i} f_{j}\displaystyle{
e^{2i\displaystyle{\sum_{l\neq i,j}\theta_{ij}(l)S^{z}_{l}}}}
b^{+}_{j\sigma}b_{i\sigma}\\
=&  \displaystyle{e^{-2i\displaystyle{\sum_{l\neq i,j}\theta_{ij}(l)
S^{z}_{l}}}}f^{+}_{i}f_{j}\tilde{b}^{+}_{j\sigma}\tilde{b}_{i\sigma}
\end{array}\end{equation}
where $\theta_{ij}(l)=\theta_{i}(l)-\theta_{j}(l)$, $\theta_{i}(l)$ is an 
angle between the direction from site $i$ to site $l$ and some fixed 
direction, the $x$ axis for example; $S^{z}_{l}=\frac{1}{2}b^{+}_{l\alpha}
\sigma^{z}_{\alpha\beta}b_{l\beta}$, the $z$-component of the spin 
operator; $\tilde{b}_{i\sigma}=e^{2i\sum_{l\neq 
i}\theta_{i}(l)S^{z}_{l}}b_{i\sigma}$, is a fermion operator. 
Under the approximations (8) and (9), and eliminated the magnetization 
density field $\hat{L}(x)$, the Lagrangians in (4) and (5) can be written 
as, respectively
\begin{equation}
{\it L}_{\Omega}=\frac{1}{2g_{0}}\int 
d^{2}x[(\vec{\partial}\hat{\Omega})^{2}+\frac{1}{c^{2}}(\partial_{\tau}
\hat{\Omega})^{2}]
\end{equation}
\begin{equation}
{\it L}_{\xi}= \displaystyle{\sum_{i}}\xi^{*}_{i}(\partial_{\tau}-\mu_{i})
\xi_{i}
+  \sqrt{2}tS\displaystyle{\sum_{<ij>}\{\xi^{*}_{j}\xi_{i}
e^{i\gamma_{ij}^{'}}[1+\eta_{i}\eta_{j}\hat{\Omega}(x_{i})\hat{\Omega}
(x_{j})]^{\frac{1}{2}}}+h.c\}
\end{equation}
where $\gamma^{'}_{ij}=\gamma_{ij}+\displaystyle{\sum_{l\neq i,j}}
\theta_{ij}(l)_{S}<\hat{\Omega}|(b^{+}_{l\uparrow}b_{l\uparrow}-
b^{+}_{l\downarrow}b_{l\downarrow})|\hat{\Omega}>_{S}$, 
$g_{0}=(J(1-\delta)^{2}S^{2})^{-1}$, $c^{2}=8(aJ(1-\delta)S)^{2}$. For the 
$J$-term in (4), we have replaced the $f^{+}_{i}f_{i}$ and 
$f^{+}_{j}f_{j}$ by $\delta=<f^{+}_{i}f_{i}>=<f^{+}_{j}f_{j}>$, the 
doping density. We have omitted the terms $\sum_{i}\eta_{i}\omega(x_{i})$ 
and $\sum_{i}\eta_{i}\omega(x_{i})\xi^{*}_{i}\xi_{i}$. If $\omega(x)$ is 
a slowly varying function of space coordinates $\vec{x}$ and "time" 
$\tau$ and the occupation number of the quasiparticle $\xi$ is equal at 
the even and odd sites, these two terms have a little contribution to the 
system. However, the quantity $\omega(x)$ provides an attractive 
interaction between the fermions $\xi_{i}$ and $\xi_{i+\hat{\delta}}$, 
$\hat{\delta}=(\pm a, \pm a)$, at the even and odd sites, respectively,
which may induce the pairing between the slave fermions at the even and odd 
sites. Here we assume this effect is very small, and do not consider it, 
or we only consider the normal state of the system. 

For the strongly antiferromagnetic correlation among the spin degrees of 
the system,
taking the Hartree-Fock approximation, the Lagrangian (12) can be written as
\begin{equation}\begin{array}{rl}
{\it L}_{\xi}=&  \displaystyle{\sum_{i}}\xi^{*}_{i}(\partial_{\tau}-
\mu_{i})\xi_{i}+\sqrt{2}tS\chi\displaystyle{\sum_{<ij>}\xi^{*}_{j}
\xi_{i}e^{i\gamma^{'}}}\\
+&  
2atS\eta\displaystyle{\sum_{i}\xi^{*}_{i}\xi_{i}|\vec{\partial}
\hat{\Omega}(x_{i})|}
\label{a}\end{array}\end{equation}
where $\chi=<[1+\eta_{i}\eta_{j}\hat{\Omega}(x_{i})\cdot\hat{\Omega}
(x_{j})]^{1/2}>,\;\; \eta=<e^{i\gamma_{ij}^{'}}>, \;\; |\vec{\partial}
\hat{\Omega}|\equiv|\partial_{x}\hat{\Omega}|+|\partial_{y}
\hat{\Omega}|$. We have omitted the fluctuation phase of the fields $\chi$ 
and $\eta$, and taken them as constants.
The effective Hamiltonian for the charge part can be written as 
\begin{equation}
H_{\xi}=\bar{t}\sum_{<ij>}\xi^{+}_{j}\xi_{i}e^{i\gamma_{ij}^{'}}+
V\sum_{i}\xi^{+}_{i}\xi_{i}|\vec{\partial}\hat{\Omega}(x_{i})|
\label{b}\end{equation}
where $\bar{t}=\sqrt{2}tS\xi, \; V=2atS\eta$. Because of the strongly 
antiferromagnetic correlation among the spin degrees, the spin parameter 
field $\hat{\Omega}(x)$ is slowly varying in the coordinate space, so the 
phase factor $\gamma_{ij}$ is very small and can be omitted, the phase 
factor $\gamma_{ij}^{'}$ is
\begin{equation}
\gamma_{ij}^{'}=\sum_{l\neq i,j}\theta_{ij}(l)\;_{S}\!<\hat{\Omega}|
b^{+}_{l\uparrow}b_{l\uparrow}-b^{+}_{l\downarrow}b_{l\downarrow}
|\hat{\Omega}>_{S}
\label{c}\end{equation}
which is a rapid varying quantity of the lattice sites, so generally, we 
cannot treat it in the continuous limit. 
Here we omit a gauge field $\vec{A}$ which describes the interaction 
between the spin and charge degree parts of the electrons. Under the 
spin-hole coherent state representation, if taking the long range 
antiferromagnetic N\`{e}el order as a background of the spin degree part, 
the charge and spin degree parts are only coupled via the rapidly varying 
phase factor $\gamma^{'}_{ij}=\gamma_{ij}+\sum_{l\neq 
i,j}\theta_{ij}(l)\;_{S}\!<\hat{\Omega}|2S_{z}|\hat{\Omega}>_{S}$, 
$\gamma_{ij}=(\vec{x}_{i}-\vec{x}_{j})\cdot\vec{A}(\frac{x_{i}-x_{j}}{2})$
being contributed from the localized spin degree part. 
If the phase factor $\gamma_{ij}$ is a smooth varying function in 
coordinate space, this gauge field $\vec{A}$ must be massive, because the 
current corresponding to $\vec{A}$ must be conserved, there appears a 
term $<\xi^{+}_{j}\xi_{i}|\vec{\partial}\hat{\Omega}|>\cdot 
e^{i\gamma^{'}_{ij}}$ in equations (13) and (14), which provides a massive 
term to $\vec{A}$. On the 
other hand, it is reasonable to omit this gauge field $\vec{A}$ that for 
strongly antiferromagnetic correlation among the localized spin degrees, the 
phace factor $\sum_{l\neq 
i,j}\theta_{ij}(l)\;_{S}\!<\hat{\Omega}|2S_{z}|\hat{\Omega}>_{S}$ is a 
rapidly varying function in coordinate space, so the phase factor induced 
by the gauge field $\vec{A}$ can be omitted. Then for a weakly 
antiferromagnetic correlation case, we must consider the effect produced 
by this gauge field. Just done as above, we can also adopt the slave 
boson method to deal with the t-J model, and obtain the similar 
Lagrangian as (\ref{a}) or effective Hamiltonian as (\ref{b}) only if we 
consider $\xi$ as a hard-core boson field\cite{4b}. So we consider the 
Lagrangian 
(\ref{a}) or effective Hamiltonian (\ref{b}) is valid for slave fermion and 
boson descriptions, for slave fermion description, $\xi$ is a fermion field, 
for the slave boson description, $\xi$ is a hard-core boson field.

\section{Transport property of the normal state}

Now we study the effective Hamiltonian (\ref{b}). In Ref.\cite{12a}, the 
authors have studied the effect of a strongly fluctuating gauge field on 
a degenerate hard-core Bose liquid, shown that the gauge fluctuation 
causes the boson world lines to retrace themselves, and found a transport 
relaxation rate of the order of $1/\tau_{tr}\sim 2k_{B}T$, consistent with 
the normal 
state of the cuprate superconductors. The results obtained in \cite{12a} 
are also valid for the effective Hamiltonian (\ref{b}), because the 
rapidly varying phase factor $\gamma^{'}_{ij}$ provides a strongly 
staggered magnetic field which enforces the world lines of the slave boson 
(or fermion) to retrace themselves and induces the charge degrees having 
the order of $1/\tau_{tr}\sim 2k_{B}T$ transport relaxation rate.
However, we can use this result only to explain the linear dependence of 
the resistivity $\rho$ on temperature. In order to study the temperature 
dependence of the Hall coeficient (or more important, the Hall angle), we 
must introduce an external magnetic field to the phase factor 
$\gamma^{'}_{ij}$, while because the phase factor $\gamma^{'}_{ij}$ is 
a rapid varying function of the coordinate space, we cannot treat it in 
the continuous limit. To get more valid informations, we adopt this 
scenario that we separate the rapidly varying phase factor 
$\gamma^{'}_{ij}$ into two parts 
\begin{equation}\begin{array}{rl}
\gamma^{'}_{ij}=& \gamma^{(1)}_{ij}+\gamma^{(2)}_{ij}\\
\gamma^{(1)}_{ij}=& \displaystyle{\sum_{l\neq i,j}}\theta_{ij}(l)
\;_{S}\!<\hat{\Omega}|b^{+}_{l\uparrow}b_{l\uparrow}|\hat{\Omega}>_{S}\\
\gamma^{(2)}_{ij}=& \displaystyle{-\sum_{l\neq i,j}}\theta_{ij}(l)
\;_{S}\!<\hat{\Omega}|b^{+}_{l\downarrow}b_{l\downarrow}|\hat{
\Omega}>_{S}\label{d}
\end{array}\end{equation}
and introduce three slave particles
\begin{equation}
\xi=\psi\bar{\chi}\chi, \;\;\; \xi^{+}\xi=\psi^{+}\psi
=\bar{\chi}^{+}\bar{\chi}=\chi^{+}\chi
\label{e}\end{equation}
to describe the charge degree part. $\bar{\chi}$ describes a slave 
fermion moving in a background "magnetic" field produced by the phase 
factor $\gamma^{(1)}_{ij}$, $\chi$ describes a slave fermion moving in a 
background "magnetic" field produced by the phase factor 
$\gamma^{(2)}_{ij}$, $\psi$ describes a slave boson or a slave fermion 
only responsing to external magnetic and electric fields, or more 
intuitively, it can be considered as describing the "mass-centre" of the 
slave fermions $\bar{\chi}$ and $\chi$. However, corresponding to these 
slave boson and slave fermions, there exist two gauge freedoms
\begin{equation}\begin{array}{rl}
\psi & \rightarrow\displaystyle{e^{i\theta}\psi, \;\;\; \bar{\chi}\rightarrow
e^{-i\theta}\bar{\chi}, \;\;\; \chi\rightarrow\chi}\\
\psi & \rightarrow\psi, \;\;\; \bar{\chi}\rightarrow\displaystyle{
e^{i\bar{\theta}}\bar{\chi}, \;\;\; \chi\rightarrow e^{-i\bar{\theta}}\chi}
\label{f}\end{array}\end{equation}
that introduce two gauge fields. While two current conservation equations 
corresponding to these two gauge fields and the gauge invariances will
maitain the freedom of the system being conservative.
Substituting equation (\ref{e}) into equation (\ref{b}), we have
\begin{equation}
\bar{H}=\bar{t}\displaystyle{\sum_{<ij>}\psi^{+}_{j}\psi_{i}(
\bar{\chi}^{+}_{j}\bar{\chi}_{i}e^{i\gamma^{(1)}_{ij}})(\chi^{+}_{j}
\chi_{i}e^{i\gamma^{(2)}_{ij}})}
+ V\displaystyle{\sum_{i}}\psi^{+}_{i}\psi_{i}|\vec{\partial}\hat{
\Omega}|\label{g}
\end{equation}
Under the Hartree-Fock approximation, we can have the following 
Lagrangian correponding to the Hamiltonian (\ref{g})
\begin{equation}\begin{array}{rl}
{\it L}=& \displaystyle{\sum_{i}}\{\psi^{*}_{i}(\partial_{\tau}-
\lambda_{i})\psi_{i}+\bar{\chi}^{*}_{i}(\partial_{\tau}+\lambda_{i}
+\eta_{i})\bar{chi}_{i}\\
+& \chi^{*}_{i}(\partial_{\tau}-\eta_{i})\chi_{i}\}
+V\displaystyle{\sum_{i}}\psi^{*}_{i}\psi_{i}|\vec{\partial}\hat{
\Omega}|\\
+& \bar{t}\displaystyle{\sum_{<ij>}\{A_{ij}\psi^{*}_{j}\psi_{i}
+B_{ij}e^{i\gamma^{(1)}_{ij}}\bar{\chi}^{*}_{j}\bar{\chi}_{i}
+C_{ij}e^{i\gamma^{(2)}_{ij}}\chi^{*}_{j}\chi_{i}\}}
\label{h}\end{array}\end{equation}
where, $m_{\psi}=(A\bar{t})^{-1},\; m_{\bar{\chi}}=(B\bar{t})^{-1}, \; 
m_{\chi}=(C\bar{t})^{-1}$,  
$A_{ij}=<\bar{\chi}^{+}_{j}\bar{\chi}_{i}e^{i\gamma^{(1)}_{ij}}>
<\chi^{+}_{j}\chi_{i}e^{i\gamma^{(2)}_{ij}}>=Ae^{i\Theta_{ij}}, \;
B_{ij}=<\psi^{+}_{j}\psi_{i}><\chi^{+}_{j}\chi_{i}e^{i\gamma^{(2)}_{ij}}>
=Be^{-i\Theta_{ij}+i\bar{\Theta}_{ij}}, \;
C_{ij}=<\psi^{+}_{j}\psi_{i}><\bar{\chi}^{+}_{j}\bar{\chi}_{i}
e^{i\gamma^{(1)}_{ij}}>=Ce^{-i\bar{\Theta}_{ij}}, \;
\Theta_{ij}=(\vec{x}_{i}-\vec{x}_{j})\cdot\vec{a}(\frac{x_{i}-x_{j}}{2}),\;
\bar{\Theta}_{ij}=(\vec{x}_{i}-\vec{x}_{j})\cdot\vec{\bar{a}}(\frac
{x_{i}-x_{j}}{2})$. We introduce two Lagrangian multipliers $\lambda_{i}$ 
and $\eta_{i}$ to add the constraints \ref{e} to the system. Under the 
gauge transformations 
(\ref{f}), the Lagrangian (\ref{h}) remains invariance. 

In the continuous limit, the Lagrangian (\ref{h}) can be rewritten as
\begin{equation}\begin{array}{rl}
{\it L}=& \int d^{2}x\{\psi^{*}(\partial_{\tau}-ia_{0})\psi
+\bar{\chi}^{*}(\partial_{\tau}+ia_{0}+i\bar{a}_{0})\bar{\chi}\\
+& \chi^{*}(\partial_{\tau}-i\bar{a}_{0})\chi\}
+\int d^{2}x\{\displaystyle{\frac{1}{2m_{\psi}}\psi^{*}(\vec{\partial}
-i\vec{a})^{2}\psi}\\
+& \displaystyle{\frac{1}{2m_{\bar{\chi}}}\bar{\chi}^{*}(\vec{\partial}
+i\vec{a}+i\vec{\bar{a}}+i\vec{A})^{2}\bar{\chi}
+\frac{1}{2m_{\chi}}\chi^{*}(\vec{\partial}-i\vec{\bar{a}}-i\vec{A}^{'})
^{2}\chi\}}\\
+ & V^{'}\int d^{2}x\psi^{*}\psi|\vec{\partial}\hat{\Omega}|
\label{k}\end{array}\end{equation}
where, $V^{'}=V/a^{2}, \; (\vec{x}_{i}-\vec{x}_{j})\cdot\vec{A}(\frac
{x_{i}-x_{j}}{2})=\gamma^{(1)}_{ij}, \; -(\vec{x}_{i}-\vec{x}_{j})
\cdot\vec{A}^{'}(\frac{x_{i}-x_{j}}{2})=\gamma^{(2)}_{ij}$. It is 
reasonable in the continuous limit to study the property of the 
Lagrangian \ref{h}, because the phase factors $\gamma^{(1)}_{ij}$ and 
$\gamma^{(2)}_{ij}$ are slowly varying functions, so we can introduce 
gauge fields to describe them. However, in thermodynamic limit, we have 
$<b^{+}_{l\uparrow}b_{l\uparrow}>=<b^{+}_{l\downarrow}b_{l\downarrow}>$, 
so the gauge fields $\vec{A}$ and $\vec{A}^{'}$ can be generally written 
as $\vec{A}=\vec{A}^{'}=\vec{\bar{A}}+\delta\vec{A}, \; 
\nabla\times\vec{\bar{A}}=\bar{B}=\pi(1-\delta)$, $\delta$ is the doping 
density, while the fluctuation field can be absorbed into 
$\vec{\bar{a}}$. We see that the phase factors $\gamma^{(1)}_{ij}$ and 
$\gamma^{(2)}_{ij}$ only provide uniform "magnetic" fields to the slave 
fermions $\bar{\chi}$ and $\chi$, respectively. Under these 
approximations, we can easily treat the Lagrangian (\ref{k}).

First we show that the gauge field $\vec{a}$ is massive and the gauge 
field $\vec{\bar{a}}$ enforces the slave fermions $\bar{\chi}$ and $\chi$ 
to be confined. To do so, we consider the current-current correlations of 
the slave fermions $\bar{\chi}$ and $\chi$. Because of appearance of the 
uniform magnetic field $\bar{B}$ in the slave fermions $\bar{\chi}$ and 
$\chi$ systems, there exists a zero-field Hall conductance dynamically 
produced by this field $\bar{B}$ in their current-current 
correlations\cite{12b}\cite{12c}, 
so their current-current correlations can be generally written as
\begin{equation}\begin{array}{rl}
\Pi_{\chi\alpha\beta}=& \Pi_{\chi\bot}\displaystyle{(\delta_{\alpha\beta}-
\frac{k_{\alpha}k_{\beta}}{k^{2}})+\Pi_{\chi\|}\frac{k_{\alpha}k_{\beta}}{
k^{2}}+i\epsilon_{\alpha\beta}\omega\sigma_{xy}}\\
\Pi_{\bar{\chi}\alpha\beta}=& \Pi_{\bar{\chi}\bot}\displaystyle{(
\delta_{\alpha\beta}-\frac{k_{\alpha}k_{\beta}}{k^{2}})+
\Pi_{\bar{\chi}\|}\frac{k_{\alpha}k_{\beta}}{k^{2}}-i\epsilon_{
\alpha\beta}\omega\sigma_{xy}}\label{l}
\end{array}\end{equation}
where $\sigma_{xy}$, Hall conductance, is a constant. 
$\Pi_{\chi\alpha\beta}$ has opposite sign Hall conductance against 
$\Pi_{\bar{\chi}\alpha\beta}$ because the slave fermion $\chi$ carries 
negative charge to $\vec{A}$ while the slave fermion $\bar{\chi}$ carries 
positive charge to $\vec{A}$. In the low energy and long wavelength 
limit, $\Pi_{a\bot}$ and $\Pi_{a\|}, \;a=\chi, \bar{\chi}$, are the 
quadratic functions of $\omega$ and $k$\cite{8}.
Generally, they can be written as
\begin{equation}
\Pi_{a\bot}=\eta_{a}k^{2}-\varepsilon_{a}\omega^{2},\;\;\;
\Pi_{a\|}=\bar{\eta}_{a}k^{2}-\bar{\varepsilon}_{a}\omega^{2}.
\label{la}\end{equation}
where $\eta_{a}, \bar{\eta}_{a}, 
\varepsilon_{a}\;and\;\bar{\varepsilon}_{a}$ are constants. For the gauge 
field $\vec{a}$, after integrating out the slave fermions $\chi, 
\;\bar{\chi}$, and gauge field $\vec{\bar{a}}$, its propagator is
\begin{equation}
D^{-1}=\Pi_{\chi}(\Pi_{\chi}+\Pi_{\bar{\chi}})^{-1}\Pi_{\bar{\chi}}
\label{lb}\end{equation}
We see that, in the long wavelength limit $k\rightarrow 0$, the Hall 
conductance terms in (\ref{lb}) produce a mass term for the gauge field 
$\vec{a}$, so the gauge field $\vec{a}$ has a little influence on the 
system although the slave boson (or fermion) $\psi$ dynamically produces
an unusual term $\frac{i\omega}{k}$, we can omit it in equation (\ref{k}). 
However, the propagator of the gauge field $\vec{\bar{a}}$ reads
\begin{equation}
\bar{D}^{-1}=\Pi_{\chi}+\Pi_{\bar{\chi}}\label{lc}
\end{equation}
the Hall conductance terms in (\ref{lc}) are cancelled. After integrating 
out the slave fermions $\chi$ and $\bar{\chi}$, we obtain an effective 
action of the gauge field $\vec{\bar{a}}$ as taking a suitable scalling 
for "time" $\tau$
\begin{equation}
S[\vec{\bar{a}}]=\frac{1}{4g^{2}}\int d^{3}xF^{2}_{\mu\nu}, \;\;\;
\frac{1}{g^{2}}\sim\frac{1}{\sqrt{\delta}}.\label{ld}
\end{equation}
Here for simplicity we include the $\bar{a}_{0}$ term. If we consider the 
topologically nontrivial hedgehog configurations of the gauge field 
$\vec{\bar{a}}$ with integer topological charge $q=\frac{1}{2\pi}\int 
ds_{\mu}\epsilon_{\mu\nu\lambda}\partial_{\nu}\bar{A}_{\lambda}$, the 
confinement length of the slave fermions $\chi$ and $\bar{\chi}$ 
is\cite{12c}\cite{12d}\cite{12e}
\begin{equation}
\xi=\frac{ag}{2\pi}e^{const./g^{2}}\label{le}
\end{equation}
where $a$ is an in-plane lattice constant. However, we have two basic 
length parameters, the confinement length $\xi$ and the Landau length 
$l_{B}\propto\frac{1}{\sqrt{\bar{B}}}$. In the half filling limit, the 
confinement length of the slave fermions $\bar{\chi}$ and $\chi$ is 
determined by the Landau length $l_{B}$. On the other hand, in the 
overdoping limit, their confinement length is determined by $\xi$. 

Based upon the above discussions, the Lagrangian (\ref{k}) can be 
rewritten as \begin{equation}\begin{array}{rl}
{\it L}=& \int d^{2}x\{\psi^{*}(\partial_{\tau}-ia_{0}+iA^{ex}_{0})\psi
+\bar{\chi}^{*}(\partial_{\tau}+ia_{0}+i\bar{a}_{0})\bar{\chi}
+\chi^{*}(\partial_{\tau}-i\bar{a}_{0})\chi\\
+& \displaystyle{\frac{1}{2m_{\psi}}\psi^{*}(\vec{\partial}+i\vec{A}^{ex}
)^{2}\psi+\frac{1}{2m_{\bar{\chi}}}\bar{\chi}^{*}(\vec{\partial}+
i\vec{A})^{2}\bar{\chi}}\\
+& \displaystyle{\frac{1}{2m_{\chi}}
\chi^{*}(\vec{\partial}-i\vec{A})^{2}\chi+V^{'}
\psi^{*}\psi|\vec{\partial}\hat{\Omega}|}\}\label{lf}
\end{array}\end{equation}
where we add an external gauge fields $\vec{A}^{ex}$ and $A^{ex}_{0}$, 
and omit the gauge field $\vec{a}$ and $\vec{\bar{a}}$. Although the 
slave boson (or fermion) $\psi$ dynamically contributes a term 
$(\chi_{F}k^{2}-\frac{i\omega}{v_{F}k})(\delta_{\alpha\beta}-\frac{
k_{\alpha}k_{\beta}}{k^{2}})a_{\alpha}a_{\beta}$ to the gauge field 
$\vec{a}$, the mass term derived from the slave fermions $\bar{\chi}$ and 
$\chi$ for the gauge field $\vec{a}$ will remove the singular behavior of 
its propagator, and maitains the Fermi liquid behavior of the slave boson 
(or fermion) $\psi$ invariance. 
However, the density constraints in (\ref{e}) and the current conservation 
law of the slave particle fields $\psi,\; \chi$ and $\bar{\chi}$ enforce 
the longitudial currents to satisfy the following equation
\begin{equation}
\vec{J}_{\psi\|}=\vec{J}_{\chi\|}=\vec{J}_{\bar{\chi}\|}\label{lg}
\end{equation}
while for the transversal currents there are not any constraints. We see 
that the slave particle fields $\psi, \; \chi$ and $\bar{\chi}$ interact 
on each other only via the scalar gauge fields $a_{0}$ and $\bar{a}_{0}$. 
If we redefine the scalar gauge field, $a_{0}+\bar{a}_{0}=-a^{'}_{0}$, 
then we obtain the similar Lagrangian as that in Ref.\cite{12f}, so we 
can use their results about the calculations of relaxation rates.
The slave fermions $\bar{\chi}$ and $\chi$ have the same relaxation rate 
induced by the quasiparticle-scalar-gauge fluctuation scattering
\begin{equation}
\frac{\hbar}{\tau_{\chi}}=\frac{\hbar}{\tau_{\bar{\chi}}}
\simeq 2\eta(0)k_{B}T\label{lh}
\end{equation}
where $\eta(0)\sim 1$ is a constant, while the slave boson (or fermion) 
$\psi$ has the relaxation rate
\begin{equation}
\frac{\hbar}{\tau_{\psi}}=\eta^{'}(0)\frac{(k_{B}T)^{2}}{t}
\label{lk}\end{equation}
where $\eta^{'}(0)$ is a constant. Because in the real case, we have 
$k_{B}T/t\ll 1$, so for external electric field we have the transport 
relaxation rate $\tau_{tr}=\tau_{\chi}\simeq 2k_{B}T$, consistent with 
that one directly calculates it[26] using the effective Hamiltonian 
(14), it also shows that the separations in (16) and (17) are 
reasonable. We see that the scalar gauge fields 
$a_{0}$ and $\bar{a}_{0}$ do not change the Fermi liquid behavior of the 
slave boson (or fermion) $\psi$, so we find that the charge degree part 
described by the Lagrangian (\ref{a}) or effective Hamiltonain (\ref{b}) has 
two relaxation rates corresponding to different responses to external 
magnetic and electric fields, respectively. Here we must give a detail 
explanation about the equations (30) and (31). First we only turn on an 
external electric field, so we have gauge fields $\vec{A}^{ex}_{\|}$ and 
$A^{ex}_{0}$. If we take a gauge transformation to the slave boson (or 
fermion) $\psi$, we can cancel the gauge field $\vec{A}^{ex}_{\|}$, and 
obtain an effective scalar gauge field $\bar{A}^{ex}_{0}$, so the 
response of the external electric is only the density-density 
correlations of the slave particle fields $\psi, \; \bar{\chi}$ and 
$\chi$. Because of the constraints (\ref{e}) and (\ref{lg}), there exist 
strongly interactions among the slave particles $\psi, \; \bar{\chi}$ and 
$\chi$ via the gauge fields $a_{0}$ and $\bar{a}_{0}$, which will 
drastically change this response of the external electric field. In the 
normal state, the resistivity of the system is
\begin{equation}
\rho(T)\propto 2k_{B}T(1+O(\frac{k_{B}T}{t}))+\gamma n_{i}
\label{Ak}\end{equation}
where the last term in bracket is very small $k_{B}T/t\ll 1$, $\gamma$ is 
a constantn, $n_{i}$ is the density of impurity, the last term derives 
from the impurity scattering. However, if 
we only switch on an external magnetic field, we have a gauge field 
$\vec{A}^{ex}_{\bot}$, the response of the external magnetic field is 
only the current-current correlation of the slave boson (or fermion) 
$\psi$. Although there exist strongly interactions among the slave 
particles $\psi, \; \bar{\chi}$ and $\chi$ via the scalar gauge fields 
$a_{o}$ and $\bar{a}_{0}$, the Fermi behavior of the slave boson (or 
fermion) $\psi$ is not destroyed by these scalar gauge interactions, 
so for the external magnetic field, the charge degree part only show its 
Fermi liquid behavior because only the slave boson (or fermion) $\psi$ 
response for the external magnetic field, the Hall angle of the system is
\begin{equation}
cot\theta_{H}=\frac{\rho^{\psi}_{xx}}{\rho^{\psi}_{xy}}
=\alpha T^{2}+\beta n_{i}\label{Bl}
\end{equation}
where $\alpha(\propto\frac{1}{B})$ and $\beta$ are constants, the last term 
derives from the impurity scattering. 
According to the above discussions, the anomalous transverse 
magnetoresistance is closely related to the temperature dependence of the 
Hall angle, they are derived from the same origin, the slave boson (or 
fermion) $\psi$ system. Because of the slave boson (or fermion) $\psi$ 
system remains the Fermi liquid behavior, according to the Kohler's rule 
we should have a temperature dependence of the 
magnetoresistance of the $\psi$ system
$\Delta\rho^{\psi}/\rho^{\psi}\propto(tan\theta_{H})^{2}
\propto B^{2}T^{-4}$. 
since we have the relation $\Delta\rho^{\psi}=\Delta\rho$, so we can obtain 
the 
following expression of the magnetoresistance of the charge degree part
\begin{equation}
\frac{\Delta\rho}{\rho}=\frac{\rho^{\psi}}{\rho}
\frac{\Delta\rho^{\psi}}{\rho^{\psi}}\propto B^{2}T^{-n}
\label{Cl}\end{equation}
For the resistivity $\rho\sim T$, we have $n=3$; For the resistivity
$\rho\sim T^{2}$, we have $n=4$. Generally, in the underdoping range, the 
resistivity is $\rho\sim T^{\alpha}, \; 1<\alpha\leq 2$ in the low 
temperature range, the magnetoresistance has the temperature dependence 
$\Delta\rho/\rho\propto T^{-n}, \; 3<n\leq 4$, consistent with the 
experimental data in \cite{7i}\cite{7j}. 
In the $YBa_{2}Cu_{3}O_{7-\delta}$ samples between $100$ and 
$375K$\cite{7i}, $\Delta\rho/\rho$ follows a power law $T^{-n}$, with 
$n=3.5$ and $3.9$ in the ($T_{c}=$) 90-K and 60-K crystals, respectively. In 
the 
$BSCCO\;2:2:1:2$ single-crystal samples\cite{7j}, $\Delta\rho/\rho$ is 
shown to vary as $\sim T^{-3}$ from $T_{c}$ up to room temperatures. 
For the $YBCO$ samples, there exists a $Cu-O$ chain which may affect the 
experimental results, but the changing trend of the exponential $n$, 
from the optimal doping to the underdoping cases, is consistent with the 
equation (34). We need more experimental data to testify the temperature 
dependence of the magnetoresistance given in (34).

\section{Magnetic property of the normal state}

Because the slave boson (or fermion) $\psi$ system remains the Fermi 
liquid behavior, after integrating out the field $\psi$, we can obtain an 
effective term provded by the interaction between the spin parameter 
field $\hat{\Omega}$ and the slave boson (or fermion) $\psi$
\begin{equation}
{\it L}_{\psi}[\hat{\Omega}]=-\beta\sum_{n}\int\frac{d^{2}q}{
(2\pi)^{2}}\frac{|\omega_{n}|}{\omega_{F}}|\hat{\Omega}|^{2}(q,\omega_{n})
\label{ll}\end{equation}
where $\omega_{F}\propto\frac{1}{V^{'2}k_{F}}$, a character energy scale 
describing the damping of the quasiparticle-hole pairing excitation to 
the spin wave spectrum. We must carefully pay attention on the term 
(\ref{ll}) which is not directly derived from an usual the 
quasiparticle-hole pairing excitation of a magnon, because the 
interaction term between the spin parameter field $\hat{\Omega}$ and the 
slave boson (or fermion) $\psi$ is 
$\psi^{*}\psi|\vec{\partial}\hat{\Omega}|$, a complicated interaction. 
Meanwhile, 
in the momentum space, we remain in mind that origin point of the momentum 
for the spin parameter 
field is at $\vec{q}=\vec{Q}=(\pm\pi/a,\pm\pi/a)$, while origin point of 
the momentum for the slave field $\psi$ is at $\vec{q}=(0,0)$.

From equations (11) and (\ref{ll}), we obtain an effective action of the 
spin parameter field of the t-J model
\begin{equation}
S_{eff.}[\hat{\Omega}]=\beta\sum_{n}\int\frac{d^{2}q}{(2\pi)^{2}}\{
\frac{1}{2g_{0}}(q^{2}+\frac{1}{c^{2}}\omega^{2}_{n})-\frac{|
\omega_{n}|}{\omega_{F}}\}|\hat{\Omega}|^{2}(q,\omega_{n})
\label{ka}\end{equation}
where $|\hat{\Omega}(x,\tau)|=1$, the origin points of $\vec{q}$ are in the 
corner 
points $\vec{Q}=(\pm\frac{\pi}{a},\pm\frac{\pi}{a})$. The action (\ref{ka})
is the same as that in Refs.\cite{4}\cite{4a} that we 
obtained from a p-d model or an effective Hamiltonian derived from a 
three-band Hubbard model. This action has two critical regions: one is a 
$z=1$ (where $z$ is a dynamic exponent) region which is consisted of 
three regimes: a renormalized classical (RC) regime, a quantum critical 
(QC) regime and a quantum disorder (QD) regime\cite{1}; another one is a 
$z=2$ region which maybe is also divided into the same two (QC and QD) 
regimes as above, but their behavior is completely different from that in 
the $z=1$ region. In the undoping case, $\omega_{F}\rightarrow\infty$, 
the system is in the RC regime\cite{1}\cite{2}. In the underdoping case, 
$\omega_{c}<\omega_{F}<\infty$, the system is in the $z=1$ QC and/or QD 
regimes[3][4]. In the optimal doping case, $\omega_{F}<\omega_{c}$, 
the system goes into the $z=2$ region\cite{3}\cite{4}\cite{13}. $\omega_{c}$ 
is a 
characteristic energy scale which indicases a crossover of the system 
from the $z=1$ region to the $z=2$ region as doping.
We see that the $\omega_{F}$ term in (\ref{ka}) which derives from the 
damping of the quasiparticle-hole pairing excitation to the spin wave 
spectrum is very important for determining the doping influence on the 
system, especially in the optimal doping case, this term is dominant.

Generally, in the $z=1$ region, the $\omega_{F}$ term is very small, and 
can be treated perturbatively, in the low energy limit we can obtain 
following spin susceptibility
\begin{equation}
\chi(q,\omega)=\frac{\chi_{0}}{\xi^{-2}+q^{2}-\frac{1}{c^{2}}\omega^{2}
-\frac{i\omega}{\omega^{R}_{F}}}\label{kb}
\end{equation}
where $\xi$ is a coherent length, $\omega^{R}_{F}$ is a renormalized 
characteristic energy scale of the spin fluctuation. In the ($z=1$) QC 
regime\cite{4}\cite{4a}, $\xi\sim\frac{1}{T}, 
\omega^{R}_{F}\sim\frac{\omega_{F}(\hat{l})}{T}$; In 
the ($z=1$) QD regime, $\xi$ and $\omega^{R}_{F}$ take constants. In the 
$z=2$ region, the $\omega_{F}$ term is dominant, the $\omega^{2}$ term is 
irrelevant and can be omitted, in the low energy limit we can obtain 
following spin susceptibility
\begin{equation}
\bar{\chi}(q,\omega)=\frac{\bar{\chi}_{0}}{\bar{\xi}^{-2}+q^{2}-
\frac{i\omega}{\bar{\omega}_{F}}}\label{kc}
\end{equation}
where $\bar{\omega}_{F}=\frac{\omega_{F}}{2g_{0}}$ is a renormalization 
group invariant quantity. In the ($z=2$) QC regime\cite{4}\cite{4a}, 
$\bar{\xi}^{2}\sim\frac{1}{T}$. Using these spin susceptibilities in 
(\ref{kb}) 
and (\ref{kc}), we can betterly explain the current experimental 
data\cite{5}-\cite{7b} 
of the nuclear magnetic resonance spin-lattice relaxation rate and the 
spin echo decay rate about the copper spin. 
The NMR spin lattice relaxation rate
$T_{1}$ and the spin echo decay rate $T_{2G}$ can be written as
\begin{equation}\begin{array}{rl}
\displaystyle{\frac{1}{T_{1}T}}& \propto\displaystyle
{\lim_{\omega\rightarrow 0}}\int d^{2}q|A(q)|^{2}
\frac{\chi^{\prime\prime}(q,\omega)}{\omega}
\propto\displaystyle{\frac{\xi^{2}_{i}}{\omega_{i}}}\\
\displaystyle{\frac{1}{T_{2G}}}& 
\propto[\int d^{2}qf(q)\chi^{\prime 2}(q,0)]^{1/2}
\propto\xi_{i}
\end{array}\end{equation}
where, $\omega_{i}=\omega^{R}_{F}$ (for z=1) or $\bar{\omega}_{F}$ (for z=2),
$\xi_{i}=\xi$ (for z=1) or $\bar{\xi}$ (for z=2), $A(q)\sim A$ is the 
hyperfine
coupling constant and $f(q)\sim f$ is the form factor originating
from the hyperfine interaction between the nuclear spin and the surrounding
electron spins.
In the QD regime we have
\begin{equation}\begin{array}{rl}
\displaystyle
\frac{1}{T_{1}T}& \propto\left\{ 
\begin{array}{ll}\displaystyle{\frac{\xi}{\omega_{F}(\hat{l})}},
&\mbox{z=1}\\
\displaystyle{\frac{\bar{\xi}^{2}}{\bar{\omega}}},
&\mbox{z=2}\end{array}\right.\\
\displaystyle{\frac{1}{T_{2G}}}& =const.
\end{array}\end{equation}
Similarly, in the QC regime we have
\begin{equation}\begin{array}{rl}
\displaystyle
\frac{1}{T_{1}T}& \propto\left\{ 
\begin{array}{ll}\frac{1}{\omega_{F}(\hat{l})T},
&\mbox{z=1}\\
\frac{1}{T{\bar{\omega}}^{2}_{F}},
&\mbox{z=2}\end{array}\right.\\

\displaystyle{\frac{1}{T_{2}}}& \propto\left\{ 
\begin{array}{ll}\frac{1}{T},
&\mbox{z=1}\\
\frac{1}{{\bar{\omega}}^{1/2}_{F}T^{1/2}},
&\mbox{z=2}\end{array}\right.
\end{array}\end{equation}
We see that the spin lattice relaxation rate $T_{1}$ is considerably affected
by the doping because of the quantity $\omega_{F}(\hat{l})\sim 1/k_{F}$, 
while the spin echo decay rate 
$T_{2G}$ depends upon doping
through the correlation length $\xi$. 
For the spin lattice relaxation rate of the oxygen spin, we need more 
explanation, because in the t-J model the 
spin degree of the planar oxygen,  
composed a Zhang-Rice spin-singlet with the localized planar copper spin,
is completely suppressed.
The slave fermion (or boson) operator $f_{i}$ in (1) really expresses a 
Zhang-Rice 
spin-singlet, if there only exists a commensurate strongly short-range 
antiferromagnetic correlation for the localized planar copper spins, at 
least it is true for the $YBCO$ samples, the planar oxygen spin will be 
not influenced by this commensurate antiferromagnetic correlation because 
the planar oxygen resides in the middle point of two nearest neighbor 
copper sites. So only the slave fermion (or boson) $f$ system, 
describeing the charge degree of electron, can influence the spin 
lattice relaxation rate of the planar oxygen spins\cite{13a}, just shown 
as in Section III, which obeys the Korringa-like rule because 
the response of the charge degree part shows the Fermi liquid behavior to 
external magnetic field. 

\section{Discussion and Conclusion}

Using the spin-hole coherent state representation, we have 
studied the normal state property of the t-J model in the usual slave 
boson and slave fermion treatment of the single occupation constraint, 
and shown that we can qualitatively explain the unusually magnetic and 
transport behaviors of the normal state of the cuprate superconducting 
materials by the t-J model. We think that the short range 
antiferromagnetic correlation induces the unusual behavior of the normal 
state of the cuprate materials, so it is a reasonable approximation that we 
take a long range antiferromagnetic 
N\'{e}el order as a background of the spin degree part of the system. 
Although the 
interaction between the charge degree and spin degree will destroy this long 
range order, but the system still has the short range antiferromagnetic 
order. In the undoping case, the system can be described by a non-linear 
$\sigma$-model (the t-J model reduces to the Heisenberg model). In the 
doping case, the interaction between the charge degree and spin degree 
provides a decay term to the non-linear $\sigma$-model, which describes 
the damping of the quasiparticle-hole pairing excitation to the spin wave 
spectrum, but this decay term is not directly derived from the 
quasiparticle-hole pairing excitation of a magnon because of the 
complicated interaction term between the spin parameter field 
$\hat{\Omega}$ and the slave boson (or fermion) field $\psi$.  
Using this effective Lagrangian \ref{ka}, we can betterly explain 
the unusually magnetic behavior of the planar copper spin of the normal 
state of the cuprate superconducting materials. For the planar oxygen 
spin, we think that its normal Korringa-like relaxation behavior is 
coming from the contribution of the slave particle $f$ in (1), 
described the Zhang-Rice spin-singlet and charge degree of electron. 
While because of there existing the strongly short-range
antiferromagnetic coreelation in the localized spin degree part, the 
charge degree part will feel a strongly staggered magnetic field as the 
doping hole hopping, this staggered magnetic field drastically influences 
the behavior of the charge degree part, and enforces it to have different 
responses to external magnetic field and electric field and to show two 
relaxation rate behaviors corresponding to the planar resistivity and 
Hall angle, respectively. This character of the charge degree part 
responsed to external magnetic field is compatible with the Korringa-like 
relaxation behavior of the planar oxygen spin. According to thses 
properties of the responses of the charge degree part to external 
magnetic and electric fields, we have calculated the 
temperature dependence of the magnetoresistance, and found that near the 
optimal doping, it varies as $T^{-n}, \; n\sim 3$, in the underdoping 
cases, it varies as $T^{-n}, \; n\sim 4$, consistent with the current 
experimental data\cite{7i}\cite{7j}. The transport relaxation rate is of 
the order of $2k_{B}T$, consistent with the normal state of the cuprate 
superconductors. Of course, the results we have obtained are invalid in 
the half doping limit, in that case the doping hole tends to localize due 
to the strong interaction with the nearest copper spin; they are also 
invalid in the overdoping limit where the antiferromagnetic correlation 
is very weak and/or there exists a transition from two-dimensional system 
to three-dimensional system, because it is not reasonable to take a long 
range antiferromagnetic N\`{e}el order as a background of the spin degree 
part and the charge and spin degrees of electron are confined.

\end{document}